\documentclass[12pt,a4paper,twoside]{article}

\usepackage{graphicx}                 
\usepackage{hyperref}                 
\usepackage[latin1]{inputenc}
\usepackage{wrapfig}
\usepackage{fixltx2e}
\usepackage{listings}
\date{\small 2011-12-06}
\title{Discovering novel computer music techniques by exploring the space
of short computer programs
}

\author{ Ville-Matias Heikkilä }

\begin{document}
\maketitle

\begin{abstract}
        Very short computer programs, sometimes consisting of as few as
        three arithmetic operations in an infinite loop, can generate data
        that sounds like music when output as raw PCM audio. The space of  
        such programs was recently explored by dozens of individuals within
        various on-line communities. This paper discusses the programs
        resulting from this exploratory work and highlights some rather
        unusual methods they use for synthesizing sound and generating 
        musical structure.
\end{abstract}

\section{Introduction}

        During the history of computer music and algorithmic composition, a
        plethora of technical and mathematical concepts has been tried out,
        ranging from simple shift registers and cellular automata to complex
        physical and statistical models. It is easy to get the impression   
        that every potentially useful simple concept has already been at
        least superficially studied.

        In order to find new, computationally simple concepts useful for
        low-complexity computer music, a rather non-systematic excursion was
        undertaken into the space of very short computer programs that
        output raw sound data.

The excursion started from a set of seven short C-language programs
that were presented in a YouTube video\cite{video1}. As the video gained
widespread attention, several individuals spontaneously contributed
their own programs, some of which were also collected in a text
file\cite{collection}. There was never any coordination of the explorative
process, although two additional YouTube videos presenting newly found
programs were produced\cite{video2}\cite{video3} and two blog
posts were written on the
subject\cite{blogpost1}\cite{blogpost2}. The analysis in this paper has
mostly been based on the latter post.

The interest shown on various on-line communities towards the
explorative process and the resulting music suggests relevance to
artistic practices such as demoscene, glitch art, chip music and 
live coding. The term ``bytebeat'' was recently coined for referring to the
type of music.

\section{Technical framework}
The short C programs that initiated the explorative process follow the form
\begin{lstlisting}
main()
{
  int t=0;
  for(;;t++) putchar(EXPRESSION);
}
\end{lstlisting}
where \emph{EXPRESSION} is where all the variation takes place.
While the
expression is typically evaluated with 32 or more bits of integer 
accuracy, only the eight lowest bits of each result show up in the
output. These bytes are to be interpreted as unsigned 8-bit PCM 
sample values with the rate of 8000 samples per second, which is the
default configuration of the usual command-line audio output methods
available on Linux and some other Unix-like operating systems, i.e.  
the \emph{/dev/dsp} device file and the \emph{aplay} utility.

        The most fruitful stage of the exploration started with the
        appearance of an on-line tool\cite{jstool} that allowed the user to enter an
        expression and immediately listen to its output. Another, slightly
        later tool\cite{astool}, based on Flash, allowed for real-time modification of
        the expression. These tools are based on JavaScript and
        ActionScript, respectively, and therefore follow the conventions of
        said languages. The expression syntax of these languages is similar
        to that of C, but the differences in functional details introduced 
        some compatibility problems.

 All the expressions featured in this paper belong to a subset of C
 expressions where function calls are not allowed and the only
 variable is the time counter t which is never modified within the
 expression. This subset is fully compatible with the on-line tools,
 and it covers a majority of the collected expressions. The following table
 summarizes the available operators in order of precedence.

\begin{tabular}{l l l l l} \label{tab:operators}
        bitwise complement, type cast & \textasciitilde & () \\
        multiplication, division, modulus & * & / & \% \\
        addition, subtraction & + & - \\
        bit shift left/right & \textless\textless & \textgreater\textgreater \\
        less/greater than (or equal to) & \textless & \textless= & \textgreater & \textgreater= \\
        (not) equal to & == & != \\
        bitwise AND & \& \\
        bitwise exclusive OR & \textasciicircum \\
        bitwise inclusive OR & \textbar \\
        ternary conditional & ? : \\
\end{tabular}

        As the expressions have a lot of bitwise operations and the usual
        mathematical notation for them is quite cumbersome, plain C
        expression syntax will be used throughout. Table xxx lists the
        operators in order of precedence.

\section{Findings}

\subsection{General properties of the expressions}

The October 2011 version of the oneliner music formula
collection\cite{collection} contains
71 formulas, 58 of which fall within the previously-defined subset of C
expressions.

Bitwise operations and bit shifts feature prominently in the collected
expressions, on all levels of music generation. In fact, the only one that
lacks these operations altogether is the trivial minimal example: t. It has
been very difficult to find prior examples of such a reliance on binary
arithmetic in music, so we may assume that binary arithmetic are often used
in novel ways in the expressions.

Many of the expressions have been discovered purely by chance. Several
contributors have admitted that they do not understand the inner workings
of their discoveries at all. The longer expressions, however, often show a 
more structured and deterministic approach. The more structured expressions
are also more likely to use** more traditional approaches than those based 
on trial-and-error discoveries.

\subsection{Bitwise operations with amplitude values}

One of the most interesting short expressions consists of a bitwise AND
between a fast-changing and a slow-changing subexpression. Expressions like

\begin{lstlisting}
        t&t>>8
\end{lstlisting}

came to be called ``Sierpinski harmonies'' because the plotted amplitude
values form a pattern that resembles a Sierpinski triangle. When listened, a
Sierpinski harmony appears to have a multitonal melody with mostly octave   
intervals.

The first subexpression in \textit{(t\&t\textgreater\textgreater 8)} is \textit{t}, which forms a
simple eight-bit approximation of a sawtooth wave. In order to understand what the bitwise
AND does to this wave, we will analyze this wave as a sum of eight square
waves, each of which corresponds to a single bit in the amplitude value:

\begin{lstlisting}
        (t&128) + (t&64) + (t&32) + ... + (t&1)
\end{lstlisting}

Normally, the harmonic contents of these square waves are interpreted by the
human brain as belonging to a single timbre. However, when a new square-wave
component is introduced abruptly, it is likely to be interpreted as a
separate tone. One can test this effect by comparing the
hearing experience of
\begin{lstlisting}
        t&96
\end{lstlisting}
to that of
\begin{lstlisting}
        t&96&t>>8
\end{lstlisting}
which contains the same combination of two square waves in the end.

The figure below demonstrates the harmonic and melodic progression of the
Sierpinski harmony \textit{t\&t\textgreater\textgreater 8}.

\begin{center}
\includegraphics[width=14cm]{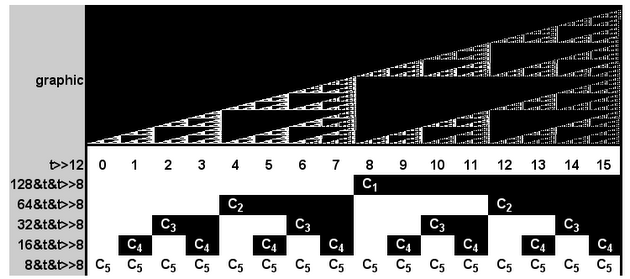}
\end{center}

If we speed up the carrier wave with a ratio that is not a power of two, we
will get non-octave intervals between the component waves. In
\begin{lstlisting}
        3*t&t>>8
\end{lstlisting}
the melody differs clearly from \textit{t\&t\textgreater\textgreater 8}. The reason for this is that the
higher-pitched square waves become dominated by their alias waves that
harmonize with the Nyquist frequency instead of the carrier.

When combining several Sierpinski harmonies with different carrier and
modulator speeds we can get more complex melodies, often sounding like
lullabies. Examples from the collection include:
\begin{lstlisting}
        t*5&t>>7|t*3&t>>8
        t*5&t>>7|t*3&t>>10
        t*9&t>>4|t*5&t>>7|t*3&t>>10
\end{lstlisting}
The expression from ``Rrrola'',
\begin{lstlisting}
        t*(0xCA98>>(t>>9&14)&15)|t>>8
\end{lstlisting}
uses bit-masking for transposing and fading out a short constructed melody
element that is encoded as as a constant.

More complex musical structures can be formed by combining several
bit-shifted modulators with bitwise operators. An example of this would be
\begin{lstlisting}
        (t>>6^t>>8|t>>12|t)&63
\end{lstlisting}
which is used as melody generator in a more complex formula that also
includes a constructed bassline and a simple drum. This expression,
by ``Mu6k'', has been analyzed in detail in \cite{blogpost2}.

The effectivity of long bit shifts for macro-level musical structure is
probably related to the effectivity of power-of-two structural ratios in pop
music. This is exemplified by concepts such as ``eight-bar blues'',
``sixteen-bar blues'' and ``thirty-two-bar form''.

\subsection{Pitch values from bitwise arithmetic}

Using a tone generator with a variable wavelength is a more traditional
method of melody synthesis. The shortest way to do this in our subset of C
expressions is by multiplying t with a subexpression that yields suitable
pitch values. A simple example, separately discovered by at least three 
individuals, is what came to be called ``The Forty-Two Melody'':
\begin{lstlisting}

        t*(42&t>>10)
\end{lstlisting}

In this case, the melody played out with the sawtooth generator \textit{t*}
is represented by a series of integers formed by zeroing out all bits except
three (42 = 101010\textsubscript{2}) in consecutive binary numbers. In the
following figure, the series is translated into familiar names of muslcal
notes.

\begin{center}
\includegraphics[width=14cm]{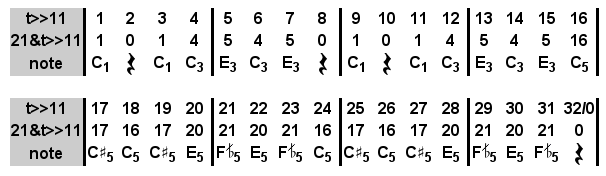}
\end{center}

Integer series with high values are likely to contain intervals that are
dissonant to Western listeners. Modulus can be used to restrict the values
into a more familiar range while retaining the effect of the higher bits: 
\begin{lstlisting}
        t*((42&t>>10)%14)
\end{lstlisting}
A series that ands with 42 (or 21) has a repetitive cycle of 32 pitch value
changes. An example of a much longer series of pitch values formed by a
simple expression is
\begin{lstlisting}
        ((t>>10)\^(t>>10)-2)%11*t&64
\end{lstlisting}
where the pitch value depends on the number of zero bits in the end of t>>10
which is used as the index to the series. The first sixteen notes are given
by the following table.

\begin{tabular}{l l l} \label{tab:rygmelody}
i & (i\textasciicircum i-2)\%11 & note \\
\hline
        0000\textsubscript{2} &  -2 &            C\textsubscript{2} \\
        0001\textsubscript{2} &  2 &             C\textsubscript{2} \\
        0010\textsubscript{2} &  6 &             G\textsubscript{3} \\
        0011\textsubscript{2} &  2 &             C\textsubscript{2} \\
        0100\textsubscript{2} &  3 &             G\textsubscript{2} \\
        0101\textsubscript{2} &  2 &             C\textsubscript{2} \\
        0110\textsubscript{2} &  6 &             G\textsubscript{3} \\
        0111\textsubscript{2} &  2 &             C\textsubscript{2} \\
        1000\textsubscript{2} &  8 &             C\textsubscript{4} \\
        1001\textsubscript{2} &  2 &             C\textsubscript{2} \\
        1010\textsubscript{2} &  6 &             G\textsubscript{3} \\
        1011\textsubscript{2} &  2 &             C\textsubscript{2} \\
        1100\textsubscript{2} &  3 &             G\textsubscript{2} \\
        1101\textsubscript{2} &  2 &             C\textsubscript{2} \\
        1110\textsubscript{2} &  6 &             G\textsubscript{3} \\
        1111\textsubscript{2} &  2 &             C\textsubscript{2}
\end{tabular}

\subsection{Modular wrap-around of amplitude values}

In ordinary software synthesis, the overflowing of an amplitude value is
generally regarded as an avoidable artifact. In many expressions in the 
formula collection, however, these artifacts form an integral part of the
resulting sound. A dramatic example would be a modification of (),
\begin{lstlisting}
        (t*9&t>>4|t*5&t>>7|t*3&t>>10)-1
\end{lstlisting}
where the simple subtraction -1 effectively introduces a percussive
instrument by turning minimum amplitude values into maximum values. As we
have not conducted spectral analysis for expressions that use modular
wrapping, we will merely give a set of expressions where the wrap-around is
used for various effects.
\begin{lstlisting}
        (t&t%255)-(t*3&t>>13&t>>6)

        (int)(t/1e7*t*t+t)%127|t>>4|t>>5|t%127+(t>>16)|t

        t>>6&1?t>>5:-t>>4

        t>>4|t&((t>>5)/(t>>7-(t>>15)&-t>>7-(t>>15)))
\end{lstlisting}
The second and fourth example divide by zero, so they will not work
directly in most implementations of C.

\section{Conclusion}

The collected expressions extensively use binary and modular arithmetic in
unusual ways we have not been able to find prior examples of. This makes it
apparent that new computational techniques for generation of sound and music
have been discovered with a simple trial-and-error methodology during the   
exploration.

The discovery process could be enhanced by various means. New formulas could
be generated automatically or semi-automatically by using genetic algorithms
or other techniques. A dedicated social website with a voting capability
would be helpful for separating interesting discoveries from less
interesting ones.

Many individuals participating in the exploration have found the C-like
infix syntax limited and cumbersome for the purpose. It is, for example,
difficult to reuse previously computed values. This has already given birth
to several software projects that combine the technical concept of
``bytebeat'' with a Forth-like RPN syntax: an iOS application called
GlitchMachine\cite{glitchmachine}, a free Python reimplementation called
libglitch\cite{libglitch}, and a
not-yet-released audiovisual virtual machine called IBNIZ. It can be assumed
that increasing the range of available programming constructs would make it 
possible to discover new, musically interesting algorithmic concepts that 
cannot be found in the expression space we have been exploring so far.

\end{document}